\documentclass[sigconf]{acmart}

\setcopyright{none}
\renewcommand\footnotetextcopyrightpermission[1]{}

\usepackage{fancyhdr}
\fancypagestyle{plain}{%
   \fancyhf{} %
   \fancyfoot[L]{}
   \fancyfoot[R]{}
}
\fancypagestyle{firstpagestyle}{%
   \fancyhf{} %
   \fancyfoot[L]{}
   \fancyfoot[R]{}
}

\usepackage{balance}
\usepackage{todonotes}
\usepackage{hyperref}

\begin{document}

\title{Towards SSH3: how HTTP/3 improves secure shells}

\author{François Michel}
\affiliation{
	\institution{UCLouvain, Belgium}
}
\email{francois.michel@uclouvain.be}

\author{Olivier Bonaventure}
\affiliation{
	\institution{UCLouvain \& WELRI, Belgium}
}
\email{olivier.bonaventure@uclouvain.be}

\begin{abstract}
  The SSH protocol was designed in the late nineties to cope with the
  security problems of the \texttt{telnet} family of protocols. It brought
  authentication and confidentiality to remote access protocols and is now
  widely used.
  Almost 30 years after the initial design, we revisit SSH in the light of
  recent protocols including QUIC, TLS 1.3 and HTTP/3.
  We propose, implement and evaluate SSH3, a protocol that provides an
  enhanced feature set without compromise compared to SSHv2.
  SSH3 leverages HTTP-based authorization mechanisms to enable new
  authentication methods in addition to the classical password-based
  and private/public key pair authentications. SSH3 users can now configure
  their remote server to be accessed through the identity
  provider of their organization or using their Google or Github account.
  Relying on HTTP/3 and the QUIC protocol, SSH3 offers UDP
  port forwarding in addition to regular TCP forwarding as well as a faster
  and secure session establishment. We implement SSH3 over \texttt{quic-go}
  and evaluate its performance.

\end{abstract}





\maketitle

\section{Introduction}\label{sec:intro}

Remote access to distant computers was one of the motivations for the creation
of the ARPANET. It was usually realized through a TCP connection
between the client and distant servers. This connection was used to
authenticate the user and then exchange commands. Several protocols use this approach
including \texttt{telnet} \cite{rfc854} and \texttt{rsh/rlogin}~\cite{rfc1282}. These
protocols suffered from various security issues~\cite{bellovin1989security}
and are now deprecated. In the mid 1990s, the Secure Shell (SSH) protocol
\cite{ylonen1996ssh} was proposed as a secure alternative. It provides
several remote services over an authenticated and encrypted channel
using a single TCP connection:
$(i)$ execution of a program, $(ii)$ access to a shell session,
$(iii)$ TCP port forwarding and $(iv)$ forwarding
of X11 graphical sessions. 
Standardized in 2006, the specification describes
the version 2.0 of the protocol, also known as SSHv2.
The design of SSHv2 is complex and spread over four
documents: the high-level architecture design~\cite{rfc4251}, the user
authentication protocol~\cite{rfc4252}, the transport layer protocol~\cite{rfc4253}
and the connection protocol~\cite{rfc4254}. SSHv2 builds
its own secure channel and provides mechanisms for user
authentication and authorization.

With the growing use of the web for critical operations such as bank transactions or
e-commerce, the need for secure communications democratized the use of
HTTPS~\cite{rfc9110} with strong security guarantees coming from
the Transport Layer Security (TLS) protocol. TLS 1.3~\cite{rfc8446} now provides
comparable security features to SSHv2 with a shorter session
establishment. A lot of effort has also been put in HTTP
to enhance user authorization to control the
access to protected web resources. This led to the design of a generic
authorization protocol for HTTP~\cite{rfc7235} allowing the use of newer
techniques such as the \textit{Bearer}~\cite{rfc6750} mechanism in addition
to the older password-based authentication~\cite{rfc2617}. This enabled the
development of advanced mechanisms such as OAuth 2.0~\cite{rfc6749}.
OAuth opened the door to the OpenID Connect
standard~\cite{sakimura2014openid} extensively used by major
Internet actors such as Google and Microsoft to provide user authentication
through their identity platforms.
All this makes HTTP authorization more flexible than SSH while
keeping applications simple. With its use for multiple
scenarios, HTTP authorization is still evolving with new schemes being
developed~\cite{ietf-httpbis-unprompted-auth-05}.

Finally, the diversity of contents available on a remote computer
has evolved. SSHv2 provides the X11 forwarding service to use graphical
programs and access visual content. However, this
service is not suited to highly dynamic content such as videos and
livestreams (e.g. a high framerate camera) available on the remote server.
An alternative is to use the TCP port forwarding feature of SSH to tunnel a
streaming application. This is however not
sufficient if low latency is needed since TCP
provides a fully reliable bytestream at the cost of added latency.
Users wanting such features have to use alternatives such as the
Remote Desktop Protocol (RDP)~\cite{rdp2022} that includes UDP-based
transport for displaying a low latency remote graphical desktop.
Furthermore, with the significant growth of QUIC traffic on the
Internet~\cite{a-look-at-quic-use},
the TCP forwarding of SSHv2 is not sufficient anymore and the lack of UDP port
forwarding may become progressively more visible in the future.

We start from the above observations and consider the following question:
``\textit{What would the design of the SSH protocol look like if it were invented
today ?}''. We revisit SSH from the ground-up and propose
SSH3, a new version of the protocol based on the strong features
of modern web protocols. Running on
top of HTTP/3, SSH3 is as easy to deploy as any HTTP application. We describe the
technical details of SSHv2 and HTTP/3 in Section~\ref{sec:background} and show how
SSH3 can provide either similar or stronger features than its predecessor in
Section~\ref{sec:ssh3}.
We present our implementation in Section~\ref{sec:implementation}
and show performance improvements of SSH3 in Section~\ref{sec:evaluation}.
We then discuss other appealing aspects of SSH3 in Section~\ref{sec:discussion}.

\section{Background}\label{sec:background}
This section covers the concepts required to explain
the design choices of SSH3. Section~\ref{sec:sshv2}
describes how SSHv2 operates above 
TCP. Section~\ref{sec:http3} presents the HTTP/3 protocol and how
modern applications can leverage its protocol stack to build
advanced and easily deployable services with only little implementation
effort.
\begin{figure}[t]
    \includegraphics{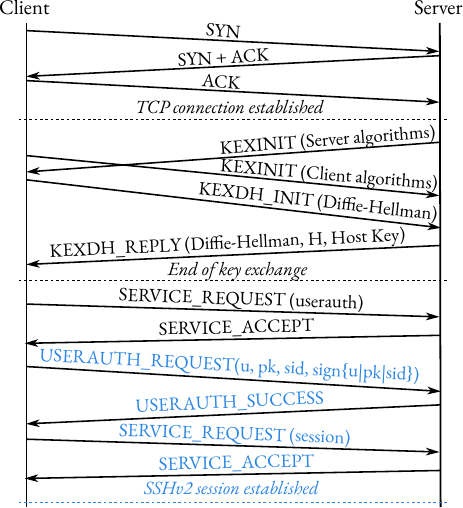}
    \caption{SSHv2 session establishment using Diffie-Hellman.
             SSH version exchange has been omitted to save space.
             Messages in blue are protected using
             the keys derived from the key exchange.
             OpenSSH uses an additional round-trip for user authentication
             by first sending the \texttt{USERAUTH\_REQUEST(none)} message to
             discover the available authentication methods.}
             \label{fig:sshv2-session-establishment}
\end{figure}
\subsection{SSHv2}\label{sec:sshv2}
Running solely atop TCP, SSHv2 defines its own mechanisms to
build a secure channel and authenticate users.
Upon successful key exchange and user authentication, users can access their
remote server and start using the services provided by SSH.
\subsubsection{Secure channel establishment}\label{sec:sshv2-secure-channel}
As they carry critical and sensitive information, the confidentiality and
authenticity of SSH communications is crucial.
The design of SSH predates TLS. For this reason, 
SSH defines its own mechanisms for establishing a secure channel to encrypt
and authenticate the session data.
\autoref{fig:sshv2-session-establishment} illustrates the establishment of an SSHv2
session. Such a session starts with a TCP handshake
depicted at the top of the Figure.
The client and server exchange \texttt{KEXINIT} messages to negotiate the algorithms
used for key exchange, encryption and hashing~\cite{rfc4253}.
Then, the actual key echange is performed
(here, using Diffie-Hellman) with the client sending the \texttt{KEXDH\_INIT} message
and the server replying with the \texttt{KEXDH\_REPLY} message. This
message also contains its host key
used for server authentication and H, a hash applied over several fields including the
negotiated Diffie-Hellman values
and the resulting shared secret. Upon reception
of these messages, both ends can derive the cryptographic keys 
to secure the connection. Additionally, a session identifier is
derived from the key exchange to uniquely identify the SSH session. This
session identifier is later used to prevent replay attacks when using
public key user authentication. The bottom of
\autoref{fig:sshv2-session-establishment} illustrates user authentication
described in the next section.

\subsubsection{User authentication}\label{sec:sshv2-userauth}
\begin{figure}[t]
    \includegraphics{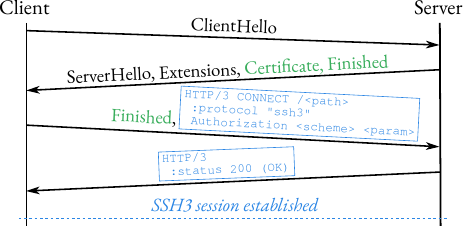}
    \caption{SSH3 session establishment. Key exchange
             is performed during the QUIC handshake.
             Messages in green
             are protected using QUIC handshake keys. Messages
             in blue are protected using the keys from the
             key exchange (1-RTT keys).
             User authentication is performed by setting the
             \texttt{Authorization} header in the HTTP/3 \texttt{CONNECT} request.
             The \texttt{<scheme>} and \texttt{<param>} variables depend on the
             used HTTP authentication scheme.}
             \label{fig:ssh3-session-establishment}
\end{figure}
The SSHv2 protocol ensures that services are provided only to authenticated and
authorized users. The SSHv2 standard defines the \texttt{user-auth}
protocol to this end~\cite{rfc4252}. In order to start the \texttt{user-auth}
protocol, the client first sends a \texttt{SERVICE\_REQUEST} message and waits
for the answer of the server.
SSHv2 supports several authentication methods, the main ones
being password-based and public key authentication. The password-based mechanism
simply consists in the client sending the username and password through the
secure channel and the server verifying that they match with the credentials on the system.
The public key authentication illustrated at the bottom of
\autoref{fig:sshv2-session-establishment} is achieved with the user performing a
signature over several fields, including
the username (u), the public key (pk) and the session
identifier (sid). The server then checks whether the public key proposed by the client
is suitable for authentication (e.g. it matches the user's public key locally
installed on the server's filesystem) and then verifies the signature.
Signing the session identifier prevents replay attacks.

Some SSH implementations such as OpenSSH or Tectia support
other ways to authenticate users. Among them is the certificate-based user
authentication: only users in possession of a certificate signed by a trusted
certificate authority (CA) can gain access to the remote server~\cite{facebook_ssh}.
Available for more than 10 years, this authentication method
requires setting up a CA and distributing the certificates
to new users and is still not commonly used nowadays.

\subsubsection{Session flow}\label{sec:sshv2-session-flow}
Once the secure channel is established and the user is successfully authenticated,
the latter can open new sessions and perform operations on the remote server.
This is done by opening SSH \textit{channels} with specific
uses such as running a shell or executing programs. Channels are
bidirectional message streams multiplexed over the single TCP connection used by
SSHv2. The messages exchanged over a channel serve several purposes. For instance,
messages can be used by a client to request the allocation of a new pseudo-terminal
(PTY), the execution of new commands or simply sending the user-typed input
to the remotely running processes. Messages can also be sent by the server to
announce the termination of the remote program along with its status or
signal code or to forward its standard and error outputs to the user.
Finally, SSH users can open channels to perform TCP port forwarding. By doing so,
the client listens on a specified local TCP port. Every connection initiated
towards this port on the client will be tunneled to the server through the SSH
channel. The server then initiates a new TCP connection towards a remote host
whose IP and port number are chosen by the client. The bytes exchanged on the TCP
connections are simply placed in new messages and sent over the channel.
Port forwarding has however its limits in SSHv2. First, as SSHv2 channels are
all multiplexed on the only TCP connection carrying the whole session,
head-of-line blocking can occur upon network losses if several TCP connections
are forwarded. Second, SSHv2 only provides TCP
forwarding and has no support for carrying UDP packets. SSHv2 is therefore unable
to tunnel UDP-based DNS, RTP or even the QUIC protocol, which now constitutes a
large part of Internet traffic nowadays~\cite{a-look-at-quic-use}.

\subsection{HTTP/3}\label{sec:http3}
During the last years, the IETF has finalized the standardization of several protocols
that provide a more secure and flexible transport than TCP. The latest
version of TLS, TLS 1.3 \cite{rfc8446} uses Diffie-Hellman and provides
perfect forward secrecy by design.
QUIC~\cite{rfc9000,langley2017quic} integrates the TLS 1.3 handshake
to provide encryption and authentication of both control and application
data as a transport feature and reduces the connection establishment duration.

The HTTP/3 protocol provides the features of HTTP/2
on top of QUIC instead
of the classical combination of TCP and TLS. Aside from the security
features discussed above, QUIC
provides modern transport features such as stream multiplexing
that allows applications to send data independently on several
streams. In addition to these byte streams, QUIC also supports
the exchange of datagrams \cite{rfc9221}. QUIC provides seamless
connection migration \cite{rfc9000} and soon multipath
communication~\cite{draft-ietf-quic-multipath},
enabling smooth network handovers that
would have disrupted the connection with TCP.

The Extended \texttt{CONNECT} HTTP extension
allows the application to directly use the underlying
QUIC stream for sending arbitrary protocol data. The application can
also open new streams and send QUIC datagrams.
This method is already extensively used by
WebTransport~\cite{ietf-webtrans-http3-08} to bootstrap new transport
connections in web browsers. 

The most appealing aspect of HTTP/3 is its native
support for user authentication~\cite{rfc9110}.
A critical part of the SSH protocol resides in the session establishment and especially
the user authentication process. HTTP already provides a solid set of mechanisms to
perform user authentication that have been implemented and used for years
for sensitive use-cases such as banking and e-commerce.

\section{SSH3}\label{sec:ssh3}

In this article, we entirely reconsider the SSH protocol stack
and propose SSH3, a modern iteration of its architecture. 
Several initiatives have already considered running SSH over the QUIC protocol,
but these propositions limit themselves to carrying classical SSH mechanisms
inside QUIC streams~\cite{quicssh,bider-ssh-quic-09}. In contrast with these
propositions, SSH3 is built above HTTP/3, not directly over
QUIC, and reconsiders the whole protocol architecture. One might wonder why
building SSH3 over HTTP/3 instead of QUIC.
An initial benefit is that HTTP/3 provides multiplexing
at the URL level: SSH3 instances can be accessible through specific
URLs. First, it allows the use of HTTP/3 proxies to act as SSH3 gateways
leading to different physical servers depending on the URL path specified in the
\texttt{CONNECT} request. Second, it can make SSH3 robust to scanning
attacks as discussed in Section~\ref{sec:scanning}.
This architecture also enables a wider
feature set than SSHv2 over TCP and reuses existing HTTP/3 mechanisms. 
The secure channel establishment is entirely performed by the QUIC layer and
the SSH3 session is initiated by sending an HTTP/3 Extended \texttt{CONNECT}
request. Furthermore, user authentication material is attached to the request
using the HTTP authorization mechanisms.
This significantly simplifies the
design and implementation aspects of SSH3 and allows implementers
to focus on the protocol features and actual system security.

\subsection{Secure connection establishment}\label{sec:ssh3-secure-channel}

\autoref{fig:ssh3-session-establishment} illustrates how an SSH3 session
is established. The first three arrows correspond to the QUIC handshake,
but the \texttt{Finished} message already carries the HTTP/3 Extended
\texttt{CONNECT} request.
With TLS 1.3, algorithms negotiation and key exchange are performed concurrently
with the client sending a \texttt{ClientHello} message and the server responding with
a \texttt{ServerHello} message, a few extensions required for key exchange and its
certificate. These operations occur during the QUIC
handshake and can be completed
during the first round-trip~\cite{rfc9001}, reducing significantly the
connection establishment time.
The first SSH3 data can already be exchanged after one
round-trip while the first round-trip was dedicated to the TCP handshake
in SSHv2, as depicted in \autoref{fig:sshv2-session-establishment}.
To derive a unique session identifier, SSH3 relies on \textit{TLS exporters}
to export cryptographic material from the QUIC session to the upper layer protocol.
Thanks to TLS exporters, both endpoints generate a same pseudorandom
value uniquely tied to the current TLS session and the running protocol.
This allows obtaining a unique session identifier similarly to SSHv2 as discussed
in Section~\ref{sec:sshv2}.
Finally, TLS natively allows authenticating both clients and servers through
certificates, similarly to the certificate-based authentication of OpenSSH.
It also works through HTTP proxies as recent HTTP mechanisms allow
forwarding client certificates through dedicated
HTTP headers~\cite{rfc9440}.
As client and server certificate-based authentication is part of TLS,
SSH3 can benefit from this
authentication method without implementation or maintenance
effort. 
The use of TLS certificates on web servers has been democratized thanks
to the efforts of Let's Encrypt~\cite{aas2019let}. Server certificates can
now be installed in a matter of seconds and provide a stronger security
than SSHv2 host keys currently used for authenticating servers.
For a private or personal use of SSH3, self-signed certificates can also be deployed,
providing a security level comparable to host keys.

\subsection{HTTP-based authentication}\label{sec:ssh3-authorization}

SSH3 uses the generic HTTP authorization
mechanism~\cite{rfc9110} and puts user authentication
material in the \texttt{Authorization} header of the \texttt{CONNECT}
request. This comes down to
setting the \texttt{<scheme>} and \texttt{<param>} variables
of \autoref{fig:ssh3-session-establishment} to appropriate values.
If the provided header suffices for authenticating and granting access
to the user, the server responds with the \textit{200 OK} HTTP response.
Otherwise, the server returns a \textit{401 Unauthorized} response.
The latter includes the \texttt{WWW-Authenticate} header indicating at
least one authentication method that can be used for subsequent requests.
Our SSH3 prototype implements three authentication techniques.

\subsubsection{Password authentication}
Password authentication can be achieved using the HTTP
\textit{Basic} authentication scheme~\cite{rfc2617}. 
In \autoref{fig:ssh3-session-establishment},
the client sets the \texttt{<scheme>} variable to \texttt{Basic}
and \texttt{<param>} to a base64-encoded value
containing the username and password. Similarly to SSHv2,
the server then compares the username and password with the local credentials
and grant access if they match.

\subsubsection{OpenID Connect authentication}\label{sec:auth-openid-connect}
The flexibility of HTTP authentication mechanisms opens the door to new
authentication methods. For instance, the OpenID Connect
protocol provides modern ways for
authenticating users using an external identity provider. This protocol
is implemented by popular Single Sign-On (SSO) services such as Google Identity
or Microsoft Entra. For instance, using this authentication method, an
SSH3 server can be configured to grant access to users that successfully logged in
to their company's identity provider using their professional email address.
The identity provider can itself implement advanced authentication methods
such as multi-factor authentication without impacting the SSH3
server configuration.
This method is entirely based on standard HTTP mechanisms such as OAUTHv2 and
the \textit{Bearer} authentication scheme~\cite{rfc6750}.
To get access to the SSH3 server, the user first requests a base64-encoded
ID token from the identity provider by logging in on its
login platform~\cite{sakimura2014openid}.
This ID token issued by the identity provider
certifies that the user successfully logged in and actually owns its
email address.
The token is passed to the SSH3 server by placing it in the
\texttt{<param>} variable of \autoref{fig:ssh3-session-establishment} and
setting the \texttt{<scheme>} variable to \texttt{Bearer}.
When decoded, the ID token is in the JSON Web Token (JWT)
format~\cite{rfc7519}.
In simple terms, JWTs can be seen as a
JSON object (payload) associated with a header and a digital signature
performed by the identity provider over the payload and the header.
The header contains descriptive
information such as the signature algorithm or the identifier of the key
used to sign the JWT. The payload contains several
fields such as a date of issue, an expiry time, the url of the identity provider
and the email address of the user.
The SSH3 server then parses the token and verifies its authenticity
using the identity provider's public key. It also checks the validity
of the JWT payload itself (e.g. ensuring the token has not expired).
Once the token has been verified and if the user's email address
is authorized by the SSH3 server, the server sends a \textit{200 OK} response
to the client's \texttt{CONNECT} request and the session can start.
ID tokens can be stored on the client (e.g. using a keyring)
and automatically renewed to avoid the user from manually logging in
at each session establishment.

\subsubsection{Public key authentication}

SSH3 also provides mechanisms for public key authentication like SSHv2.
Similarly to Section~\ref{sec:auth-openid-connect}, this method relies
on JWT tokens passed to the server using the Bearer authentication scheme.
The difference is that the JWT token is signed using the user's private key
instead of an external identity provider. The JWT token payload
contains the public key as well as the unique session identifier obtained
through the use of TLS exporters described in Section~\ref{sec:ssh3-secure-channel}.
To authenticate the user, the SSH3 server ensures that the public key in the JWT
matches the user's locally installed public key and then verifies the signature of
the JWT.
Including the session identifier in the JWT payload
provides a protection against replay attacks equivalent
to the public key authentication of SSHv2 described in
Section~\ref{sec:sshv2-userauth}. This however makes the token deviate from the
definition of a Bearer token as it does not, by design, grant the same access right
to any other party in possession of it~\cite{rfc6750}. We see this as a
minor and temporary issue. There is an ongoing standardisation effort towards
a new HTTP \textit{Signature} authentication scheme serving the exact purpose of
asymmetric key authentication. Our plan is for SSH3 to use this new authentication
scheme once it becomes standard~\cite{ietf-httpbis-unprompted-auth-05}.

\subsection{Session flow}\label{sec:ssh3-session-flow}
The SSH3 session can start once the user is authenticated. SSH3 integrates the
same concepts as SSHv2: bidirectional channels can be used to exchange
SSH messages serving different purposes: spawning new processes, carrying
process standard input and output, signalling process termination, etc.
In SSHv2, different channels are multiplexed on the same TCP connection,
leading to head-of-line blocking in case of network packet losses. With SSH3,
each channel is carried over a dedicated bidirectional QUIC stream.
Messages transmitted over a same SSH3 channel still need to be delivered
in-order to preserve the ordering of programs' output and the keystrokes
typed by the user.
Finally, SSH3 can provide both TCP and UDP port forwarding. Each TCP
connection runs over a dedicated channel.
UDP forwarding is done by tunnelling UDP packets coming in both direction
using QUIC datagrams.

\section{Implementation}\label{sec:implementation}

We implemented the different concepts presented in the article
and provide a first open source implementation of SSH3~\cite{ssh3}.
Our implementation is written in Go. The Go language
has a large network library making it easy to implement new
client and server applications on top of the HTTP semantics.
Our SSH3 implementation relies on the \texttt{quic-go}
library~\cite{quic-go}
that provides support for both the QUIC and HTTP/3 protocols.
Our implementation consists in 4000 lines
of Go code in total, including channels and messages handling,
session establishment, user authentication, unit tests as well as
the client and server command-line programs.
The implementation provides classical SSH
services such as opening interactive shells attached
to a pseudoterminal, executing single commands when interactivity
is not needed and TCP port forwarding.
It also provides the UDP port forwarding feature discussed in
Section~\ref{sec:ssh3-session-flow}.
It integrates the three authentication methods
described in this article. Depending on the user-specific configuration
on the server, users can authenticate themselves using passwords,
OpenID Connect and public key authentication. The OpenID Connect
authentication method currently opens a browser window for the user
to log in to the identity provider. This authentication method thus
currently needs a graphical environment on the client to be used.

Finally, for a smooth and progressive transition from SSHv2 to SSH3,
our prototype provides secure OpenSSH agent forwarding. For instance,
this allows users to open a shell session from their laptop to a remote
server using SSH3 and then connect from that server to legacy SSHv2
servers using the SSH keys locally
installed on the laptop. This also enables the use of SSH3
servers as gateways towards SSHv2 servers and still benefit from
SSH3 features on the first hop such as client connection migration.
\section{Evaluation}\label{sec:evaluation}
In this section, we evaluate our implementation of SSH3 through diverse
scenarios. It can be complicated to evaluate SSH3 and compare it with
SSHv2 as some SSH3 features are simply not present on SSHv2. This section
focuses on three elements. In Section~\ref{sec:eval-exec},
we show that SSH3 significantly reduces the session establishment
time compared to
SSHv2. In Section~\ref{sec:eval-typometer}, we show that our implementation
improves the user experience of an interactive shell
session. In Section~\ref{sec:eval-port-forwarding} we show that our
implementation provides port forwarding capabilities suitable for the
WAN scenario and that UDP port forwarding can be used to tunnel
UDP-based latency-sensitive applications.
Unless otherwise specified, the SSHv2 and SSH3 clients used in this
section run in machines wired to the UCLouvain university campus
network while the SSHv2 and SSH3 servers run on a server located in France
with a 15 milliseconds round-trip time towards our clients.

\subsection{Session establishment}\label{sec:eval-exec}
\begin{figure}
    \centering
    \input{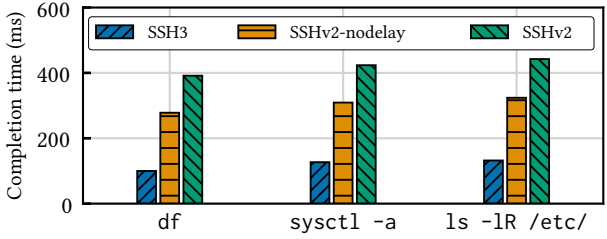}
    \caption{Completion time of non-interactive
             sessions.}
    \label{fig:session-establishment}
\end{figure}

We first evaluate the completion time of short non-interactive
sessions. We define the \textit{session completion time} as the time
needed to establish a new session, run a single command,
display the output to the user and close the session.
\autoref{fig:session-establishment} illustrates the average session
completion time for three different commands shown on the \textit{x}
axis, comparing SSH3 with the OpenSSH implementation of SSHv2.
We also modified the source code of OpenSSH to provide
\textit{SSHv2-nodelay}, a variant enabling the \texttt{TCP\_NODELAY}
socket option prior the
SSH hanshake to lower the handshake time of OpenSSH that is increased
by the Nagle algorithm~\cite{rfc1122}.
Each command is run 400 times with each solution.
We show average values since the standard deviation is low for
SSH3 (14.2) and SSHv2-nodelay (8.4) and hard to present
graphically. It is
significantly larger for the vanilla SSHv2 (more than 120 for each
command) due to the Nagle algorithm inflating the completion time.
The three commands have different output sizes on average:
582 bytes for \texttt{df}, 35kB for \texttt{sysctl} and
131kB for \texttt{ls}.  As we can see, SSH3
drastically reduces the session completion time compared to
SSHv2 and SSHv2-nodelay for every run command. This is mostly due
to the significantly shorter session establishment time discussed in
Section~\ref{sec:ssh3-secure-channel}. SSHv2
consumes additional round-trips when opening new SSH channels as
it has to negotiate channel flow control parameters
before sending data.
This can be avoided by SSH3 as it can reuse QUIC's per-stream
flow control mechanism.
Finally, the executed
command in itself has only a negligible impact on the session
completion time as they all complete rapidly and have a
relatively short output.
\subsection{Terminal responsiveness}\label{sec:eval-typometer}
\begin{figure}
    \centering
    \input{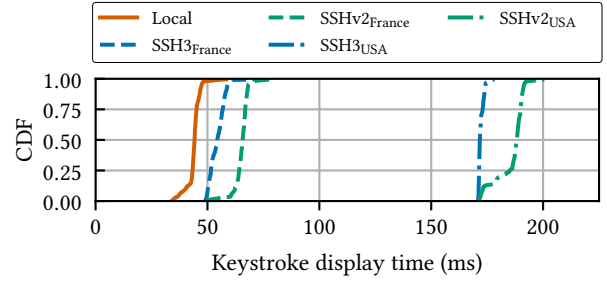}
    \caption{Keystroke latency in the Vim editor, either running
             locally (Local), over SSHv2 or over SSH3.}
             \label{fig:keystrokes-latency}
\end{figure}

We now analyze the latency experienced using
our SSH3 prototype and the OpenSSH implementation of SSHv2
in an interactive session. In this section, we want
to show that SSH3 can at least provide a latency comparable to
SSHv2. To rule out the impact of
session establishment already discussed in Section~\ref{sec:eval-exec},
we perform our experiments over established sessions.
The metric we evaluate is the \textit{keystroke latency},
computed using \texttt{typometer}~\cite{typometer}.
\texttt{typometer} records the screen while writing keystrokes in an open
text editor. The latency is the duration between the instant when a character
was typed and the instant it is actually displayed on the screen. In these
experiments, we compare the keystroke latency of the Vim text
editor~\cite{vim} opened on the remote host over SSHv2 and SSH3 sessions.
Both OpenSSH and our SSH3 implementation only display a typed character when
the server actually echoes it to the client.
This means that the keystroke latency is impacted by the network latency.
To study the impact of network latency on our scenarios, half
of the experiments are performed towards our server in France
and the other half are performed with our SSHv2 and SSH3 servers running
on a Cloudlab server located in the United States, having a 170 milliseconds
round-trip time with our clients.
\autoref{fig:keystrokes-latency} shows the keystroke latency distribution
of SSH3 and SSHv2 when an interactive session is established.
We do not experiment with our SSHv2-nodelay implementation here as OpenSSH
enables the \texttt{TCP\_NODELAY} socket option for interactive
sessions after the SSHv2 handshake.
We report under the \textit{Local} label the
keystroke latency that \texttt{typometer} recorded with Vim running
locally on the client, without SSH, as a baseline. It
therefore measures the unvoidable latency induced by the graphical desktop
environment and the text editor, also present in the SSHv2 and SSH3 curves.
As we can see on the Figure, SSH3 shows a slightly lower keystroke latency
than SSHv2. This is not surprising as the OpenSSH implementation being more
than 20 years old, it includes a series of features and complex mechanisms
inducing a non-negligeable CPU overhead. The experiments performed with the
server located in the USA help us to ensure that this difference between
SSHv2 and SSH3 is not due to the network latency as the difference
in keystroke latency between the two solutions stays similar between the
US and France while the round-trip time is tenfold.
Our prototype being significantly simpler than OpenSSH, we limit ourselves
to concluding that SSH3 can provide
a similar keystroke latency compared to SSHv2.
\subsection{TCP and UDP port forwarding}
\label{sec:eval-port-forwarding}
We also perform experiments evaluating the TCP and
UDP port forwarding features offered by SSH3.
We first study the pure throughput that can be obtained
through TCP and UDP port forwarding. We then show how real-time
media content can be accessed using the UDP port forwarding feature of SSH3.

\subsubsection{Throughput tests}
We first analyze pure throughput tests. We run the \texttt{iperf3} tool
between two \texttt{c6525-25g} Cloudlab servers
(AMD 7302P 3GHz CPUs and 25Gbps
network interfaces) with a round-trip time below 30µs.
The UDP send and receive buffers are configured with a default
size of 25MB to avoid packet loss due to buffer size.
Without any tunnelling, \texttt{iperf3} reports 22Gbps of TCP throughput
and 3.2Gbps of UDP throughput (no UDP packet loss was experienced).
The TCP throughput falls down to 4.62Gbps when \texttt{iperf3} is tunnelled
using OpenSSH port forwarding. We obtained 1.91Gbps of TCP throughput when using SSH3 TCP port
forwarding. This result is not surprising since the QUIC and HTTP/3 stacks
as well as the Go language are more computationally intensive than
the mature OpenSSH implementation. Our prototype has not been optimized for
throughput and there is room for improvement in that matter.
The obtained UDP throughput was 583Mbps when tunnelled through UDP port
forwarding, which is also significantly lower than the performance without
tunnelling. This is however sufficient for a good part of UDP utilizations,
including real-time UDP-based applications that generally favour latency
to throughput. UDP port forwarding is simply not provided by SSHv2.

\subsubsection{UDP port forwarding and real-time streaming}
We finally study the use of latency-sensitive video applications through UDP
port forwarding. We use FFmpeg~\cite{ffmpeg} to send a real-time
video stream from our client in Belgium to a
GStreamer~\cite{gstreamer} receiver
running on our server in France. The sent videos are remote
drone piloting videos recorded by Baltaci
\textit{et al.}~\cite{baltaci2022analyzing} encoded with real-time
parameters. The GStreamer receiver is configured
with a 20 milliseconds playback buffer to handle a potential small
jitter induced by QUIC pacing. Frames lost or delayed by more than 20ms will
therefore cause undesirable video artifacts on the playback. Before starting
the stream, a 5MB file is sent to probe the link bandwidth. 200 video
streams have been performed, half of it being tunnelled using the UDP port
forwarding feature of SSH3. Every received video frame was then scored using
the SSIM metric to assess its fidelity to the originally sent video frame.
In every performed experiment, every frame of every sent video has been received
in its entirety and the video has been displayed totally unaltered by the
GStreamer receiver. The sent and received videos as well as packet captures
are provided with the paper artifacts. This shows that SSH3 can be used for
accessing low-latency media resources such as camera live recordings
available on the remote server.

\section{Discussion}\label{sec:discussion}

In this section, we discuss additional use-cases for SSH3.



\subsection{Robustness to scanning and RST attacks}\label{sec:scanning}

SSHv2 is subject to port scanning attacks like many TCP-based
applications. Attackers can easily discover public SSH servers by scanning
every TCP port and finding the ones answering to SSH session establishment.
Once the attackers have discovered a public SSH endpoint, they can try
dictionary attacks on passwords.
Based over HTTP/3, SSH3 servers can avoid being publicly discovered by only
answering to SSH3 clients putting a specific \texttt{<path>} value in their
HTTP \texttt{CONNECT} request. Placing protected resources behind secret
links like this is a common behaviour in web applications. It is however
only complementary to the user authentication process.

Since SSHv2 runs above TCP, it is susceptible to attacks with spoofed TCP RST packets
\cite{arlitt2005analysis,weaver2009detecting}. Long-lived SSHv2 connections are particularly
vulnerable to such packet-injection attacks. With SSHv2, the only possibilities
to counter these attacks are to tunnel SSHv2 above IPSec or DTLS or use TCP-AO
\cite{rfc5925} or TCP MD5.
Using a secure tunnel adds operational complexity and TCP-AO relies on pre-shared keys
that are impossible to use at a large scale.
As SSH3 runs above QUIC, it is not susceptible to these attacks since all QUIC packets
and encrypted and authenticated. 

\subsection{Using SSH3 to monitor devices}

The Simple Network Management Protocol (SNMP) is often used to monitor various
types of devices. In a nutshell, the state of each device is exposed as a set of
metrics collected in a Management Information Base (MIB) which can be queried using
SNMPv3~\cite{rfc3584}. The main usage of SNMP is that
when a device receives an SNMP GET request for a specific variable, it returns its
current value.

Could SSH3 replace SNMPv3 to query MIB variables ? On many devices, the value of a MIB
variable is usually the output of a single command. For example, the \texttt{sysUpTime}
MIB variable is the uptime of the device measured in seconds. On Linux, the same
information can be obtained by using the \texttt{uptime} command. With SSH3, a
management station could simply execute this command on a remote device to query its
uptime. Since QUIC supports 0-RTT, we could obtain this information in a
single round-trip time by sending an HTTP/3 \texttt{GET} request containing
the command to execute. This would not initiate a long-lasting SSH3
session as it is done using a \texttt{CONNECT} request but would still go
through the SSH3 authentication process.
Since HTTP/3 does not provide replay protection,
we must ensure that the 0-RTT SSH3 connections are only used to carry idempotent
requests. SSH3 could restrict the utilization of 0-RTT for specific users
that only execute explicitly white-listed idempotent commands. 0-RTT would
be disabled for other users and other commands. Another benefit of SSH3 from a device
monitoring viewpoint is that \texttt{syslog} messages~\cite{rfc5424} can be transported
by the SSH3 session as QUIC datagrams.


\section{Conclusion}
In this article, we presented SSH3, a new version of the SSH protocol
rethinking its design with the modern features of the HTTP/3
protocol stack in mind. SSH3 relieves the
design and implementation complexity of the protocol by reusing the secure
mechanisms of TLS 1.3 and standard HTTP authentication mechanisms.
We showed that it significantly reduces the connection establishment time
compared to SSHv2, natively provides flexible and new ways to authenticate
users and provides new features such as UDP port forwarding and connection
migration with no compromise on the features proposed by its predecessor.
This article is a first step towards SSH3. Rethinking SSH requires
the feedback and support from the
community. We therefore encourage thoughtful comments and collaborations
to move forward and come up with a design document
describing the protocol details of SSH3. We also plan on
studying how SSH3 can be integrated to existing SSH implementations while
maintaining and improving our prototype in the meantime. Our prototype
and artefacts are publicly available~\cite{ssh3}.
\begin{acks}
   We warmly thank Thomas Wirtgen for executing the \texttt{typometer}
   experiments on his computer.
   We also sincerely thank Jonathan Hoyland, Lucas Pardue and Olivier Pereira
   for reading and providing thoughtful comments on the article.
   We thank Marten Seemann and the contributors of
   \texttt{quic-go} for its API and HTTP abstractions
   making it easy and fast to implement our SSH3 prototype.
\end{acks}

{ \balance
{
	\bibliographystyle{ACM-Reference-Format}
	\bibliography{bibliography}
}
}

\end{document}